# Crystal chemistry and *ab initio* prediction of ultra-hard rhombohedral $B_2N_2$ and $BC_2N$

Samir F. Matar[1,§,*], Vladimir L. Solozhenko[2]

[1] Lebanese German University (LGU), Sahel Alma, Jounieh, Lebanon.

https://orcid.org/0000-0001-5419-358X

[2] LSPM–CNRS, Université Sorbonne Paris Nord, 93430 Villetaneuse, France.

https://orcid.org/0000-0002-0881-9761

[§] *Former DR1-CNRS senior researcher at the University of Bordeaux, ICMCB-CNRS, France.*

[*]Corresponding author email: s.matar@lgu.edu.lb and abouliess@gmail.com

## Abstract

*New ultra-hard rhombohedral $B_2N_2$ and $BC_2N$ – or hexagonal $B_6N_6$ and $B_3C_6N_3$ – are derived from 3R graphite based on crystal chemistry rationale schematizing a mechanism for 2D → 3D transformation. Full unconstrained geometry optimizations leading to ground state energy structures and energy derived quantities as energy-volume equation of states (EOS) were based on computations within the density functional theory (DFT) with generalized gradient approximation (GGA) for exchange-correlation (XC) effects. The new binary and ternary phases are characterized by tetrahedral stacking alike diamond, visualized with charge density representations, and illustrating ion characters. Atom averaged total energies are similar between cubic BN and rh-$B_2N_2$ on one hand, and larger stabilization of rhombohedral $BC_2N$ versus cubic and orthorhombic forms (in literature assessed from favored C-C and B-N bonding), on the other hand. The electronic band structures are characteristic of insulators with $E_{gap} \sim 5$ eV. Both phases are characterized by large bulk and shear moduli and very high hardness values i.e. $H_V(rh-B_2N_2) = 74$ GPa and $H_V(rh-BC_2N) = 87$ GPa.*

1- Introduction

Graphite is known as $2H$ hexagonal carbon, i.e. with two C layers (Fig. 1a; space group $P6_3/mmc$) as well as less encountered rhombohedral $3R$ three-layered graphite was reported by Lipson and Stokes back in 1942 with $R\bar{3}m$ N°166 space group [1]. Carbon $3R$ Structure is shown in Fig. 1b using hexagonal axes for a better representation than by using a rhombohedron. Both hexagonal and rhombohedral graphite exhibit a triangular planar coordination of $sp^2$ carbon with an angle $\phi$(C-C-C) = 120°. Regarding BN polymorphs, a rhombohedral form was reported by Kurdyumov *et al.* beside wurtzite BN [2]. Graphite and layered *h*-BN (*h* for hexagonal) also called white graphite, are soft materials used as lubricants whilst diamond and cubic boron nitride *c*-BN [3] are superhard with a smaller zero pressure bulk modulus for the latter: $B_0$ ~ 380 GPa [4]. Alike diamond, *c*-BN has the zinc-blend structure with $F\bar{4}3m$ N°216 space group [5]. The structure shown in Fig. 1c is characterized by tetrahedral $sp^3$ hybridization with an angle $\phi$(N-B-N) = 109.47°. With such properties *c*-BN was considered for cutting tools replacing synthetic diamond in the second half of last century. This promoted investigations within the B-C-N ternary system characterized by light elements with small radii leading to small interatomic distances and reduced volumes. Besides BN binary, isoelectronic ternary $BC_2N$ first synthesized by Solozhenko *et al.* [6] and examined in the same year 2001 from *ab initio* by Mattesini *et al.* [7] was found to possess large hardness magnitudes in-between diamond and *c*-BN [8]. As a matter of fact BN and $BC_2N$ have the particularity of efficiently replacing diamond in industrial applications due to their higher thermal and chemical stability. A large number of such experimental studies were backed with initial strong propositions from quantum physical calculations based on the density functional theory DFT [9, 10] used herein. Indeed DFT, since its birth more than 50 years ago, has been shown throughout its accuracy in accounting for energy related physical quantities as spectroscopic ones to be predictive of new materials [7, 11]. In the context of the search for new superhard and ultra-hard materials likely to replace diamond in applications [12], present work focuses on identifying original binary B–N and ternary B–C–N candidates for superhard phases starting from original crystal chemistry rationale followed by geometry optimization and energy derived quantities from calculations within DFT. Specifically new stable rhombohedral ultra-hard $rh$-$B_2N_2$ ($h$-$B_6N_6$) and $rh$-$BC_2N$ ($h$-$B_3C_6N_3$) are proposed and labeled as extended BN and $BC_2N$ crystal networks.

2- Computational framework

The search for the ground state structure and energy was carried out using calculations based on the DFT. The plane-wave Vienna Ab initio Simulation Package (VASP) code [13, 14] was used with the projector augmented wave (PAW) method [14, 15] for the atomic potentials with all valence states especially in regard of the light elements such as B, C and N. The exchange-correlation XC effects within DFT were considered with the generalized gradient approximation (GGA) [16]. This XC scheme was preferred with respect to the local density LDA [17] one which through its over-binding character may lead to unreasonable results such as overestimated hardness. The conjugate-gradient algorithm [18] was used in this computational scheme to relax the atoms onto the ground state. The tetrahedron method with Blöchl *et al.* corrections [19] and Methfessel-Paxton [20] scheme was applied for both geometry relaxation and total energy calculations. Brillouin-zone (BZ) integrals were approximated using a special **k**-point sampling of Monkhorst and Pack [21]. The optimization of the structural parameters was performed until the forces on the atoms were less than 0.02 eV/Å and all stress components less than 0.003 eV/Å$^3$. The calculations were converged at an energy cut-off of 500 eV for the plane-wave basis set concerning the **k**-point integration in the Brillouin zone, with a starting mesh of 6×6×6 up to 12×12×12 for best convergence and relaxation to zero strains. In the post-treatment process of the ground state electronic structures, the charge density and the electronic band structures were computed and illustrated. The mechanical properties were estimated from the energy-volume equation of state EOS [22], and contemporary theoretical models of hardness [23-25].

3- Crystal chemistry rationale for predicting original 2D and 3D phases within the B–C–N system.

*Binary compounds.*

Carbon and boron nitride in the 2*H* graphitic structure are characterized by two layers at fixed *z* coordinates of ¼ and ¾ (general positions). Oppositely, in 3*R* graphite with $R\bar{3}m$ space group, carbon is at the special twofold 2*c* (*x, x, x*) Wyckoff position – or the six-fold position 6*c* (0, 0, *z*) in hexagonal axes setting. Considering the occupation of 2*c* position with B and N forming 3*R* BN sketched in Fig. 1c, is only possible by splitting the 2*c* position into two singly occupied ones: 1*a* found in *R*3*m* space group N°160. The transformation from the space group $R\bar{3}m$ to *R*3*m* is known as the *klassengleich* (same class) transition, cf. [26] and therein cited relevant works. Based on these observations, it becomes possible to consider rhombohedral BN starting from 3*R* $R\bar{3}m$ carbon with B and N sharing the 2*c* site, and then to carry out geometry relaxation to the ground state. Indeed, the new relaxed structure (Fig. 1c) belongs to the *R*3*m* space group and shows large similarities with 3*R* graphite (Fig. 1b). The crystal parameters results are presented in Table 2a. The distance d(B-N) =1.45 Å is slightly larger than d(C-C) in graphite 3R, d(C-C) =1.42 Å [1], due to the larger atomic radius of B: r(B) = 0.85 Å while r(C) = 0.77 and r(N) = 0.75 Å. The atom averaged total energy is found to be the same between 2*H* and 3*R* graphite-like BN with -8.77 eV/at. Note that regarding energy magnitudes presented herein, they are significant only in so far that they are compared within the same chemical compounds and upon establishing energy derivatives as with the energy-volume equation of states EOS [22].

A change from planar 2D sp$^2$ 3*R* BN to 3D sp$^3$ is then operated by occupying additional 1*a* positions with B1 and N1, leading after full relaxation of the atomic positions and lattice parameters to 3D $B_2N_2$ stoichiometry. Starting from the B and N substituted carbon in 3*R* graphite, a scheme of the relevant original mechanism is shown in Fig. 1d where the red arrows represent the displacements of the original B and N atoms of 3*R* towards the new B1 and N1 symbolized by ○. The major effect is the reduction of the sp$^2$ angle φ = 120° to a much smaller magnitude ~90° which eventually relaxes after calculations down to φ = 109.4°, signature of sp$^3$ tetrahedral hybridization. The new 3D boron nitride is shown in Fig. 1e where the additional B1 and N1 are highlighted with different colors; the corresponding crystal data are given in Table 2b. Regarding the electronic total energy, the atom averaged total energy E(*rh*-$B_2N_2$) = -8.71 eV/at. was found to be less than in the graphite-like phase and equal to E(*c*-BN) calculated for the purpose of comparison.

*Ternary compound.*

Accounting for a ternary phase, specifically $BC_2N$, the optimal atomic arrangement following the above schematic is somehow more complex. Expectedly, making 3D compounds between BN and carbon for applications is likely to lead to ternary phases with an intermediate stability between diamond and cubic boron nitride. In *c*-BN (Fig. 1c) as well as in hexagonal $B_6N_6$ (Fig. 1e) the structure is built up by *BN3* and *NB3* tetrahedra formed by B-N bonds, and one needs to preserve as many as possible of such bonds besides the new ones which will imply C-N and B-C. Former investigations on the search for superhard $BC_2N$ were started from *c*-BN in which such considerations were accounted for to propose orthorhombic $BC_2N$ [7]. Following the scheme presented in Fig. 1d and the resulting *rh*-$B_2N_2$, the optimal lowest energy was obtained by a succession of C2—N—B—C1—(C2—N—B—C1—C2) which can be observed running from the top of Fig. 1f depicting hexagonal $B_3C_6N_3$ after full geometry optimization. The corresponding tetrahedra are of four types: *C1(C2₃B)*, *B(N₃C1)*, *N(B₃C2)*, and *C2(C1₃N)*. As shown in Table 2c, the structure is characterized by d(C2-N) = 1.52 Å, d(C1-C2) = 1.55 Å, d(B-N) = 1.57 Å, and d(C1-B) = 1.63 Å. The distance magnitudes, compared to the distances within hexagonal $B_6N_6$ (Table 2b) show close magnitudes, except for d(B-C) = 1.63 Å, whence the expectation of a similar compactness to the binary BN 3D phases. Also for the sake of completeness, the stability of *rh*-$BC_2N$ (*h*-$B_3C_6N_3$) proposed herein was confronted with *c*-$BC_2N$ space group $P1$ (N°1) (no symmetry) [31] and *o*-$BC_2N$ space group $Pmm2$ (N°25) [7], both calculated with the same conditions of high BZ integration, namely 12×12×12 mesh. The energy results compiled in Table 1 (last three lines), show that present configuration has the lowest energy (E/at).

4- Charge density and electronic structures.

The presence of different chemical species in B,C,N involving different electronegativities ($\chi B=2.04$, $\chi C=2.55$, and $\chi N=3.04$) leads to control the bond covalence magnitude between two atoms 1 and 2 from the difference $\Delta\chi = \chi 2-\chi 1$. The resulting $\Delta\chi$ are: $\Delta\chi$(C-C)= 0.0, $\Delta\chi$(C-N)= 0.49, $\Delta\chi$(B-C)= 0.51, and $\Delta\chi$(B-N)= 1.0. These magnitudes lead to hierarchies ranging from strongly covalent for C-C bonds (as in diamond), covalent < C-N and B-C (present in $BC_2N$) <polar covalent, and Polar Covalent characterizing B-N bond in BN where one expects a charge transfer B→N resulting in charge-poor B and charge-rich N as illustrated in next paragraph.

*Charge density*.

The electronic and crystal structure relationship is further illustrated with the charge density projections onto the chemical constituents resulting from the self-consistent calculations and analyzed using VESTA software [27]. Fig. 2a shows the charge density grey volumes in *c*-BN centered on N and taking the shape of tetrahedra pointing towards 4 surrounding B as expected from the $sp^3$ hybridization. In Fig. 2b, the tetrahedral patterns of charge densities around nitrogen are reproduced in *h*-$B_6N_6$ along the hexagonal *c*-direction, but less regular patterns of charge densities are observed in the ternary *h*-$B_3C_6N_3$ (Fig. 2d) where several natures of chemical bonds are present as stated above. The most regular tetrahedral $sp^3$ shape is observed around C1 and distorted shapes around C2 as well as around N which exhibits the largest charge density as expected from the highest electronegativity of nitrogen $\chi N= 3.04$ discussed above. Then *h*-$B_3C_6N_3$ is less polar-covalent than *h*-$B_6N_6$.

*Electronic structures.*

Figure 3 shows the electronic band structures along the main directions of the Brillouin zones first wedge in face centered cubic (a) for *c*-BN, and rhombohedral (b, c) for *rh*-B$_2$N$_2$ and *rh*-BC$_2$N. All three phases possess large band gaps separating the filled valence band (VB) from the empty conduction band (CB). The zero of energy along the *y*-axis is with respect to the top of the valence band (VB), E$_V$. The band gaps are all indirect and amount to E$_{gap}$~5 eV, signaling insulating behavior. The number of bands is proportional to the number of valence states: in face centered cubic *c*-BN there are 4 bands corresponding to the 8 valence electrons, i.e. with each band containing 2 electrons. The double number of bands is found in *rh*-B$_2$N$_2$ and *rh*-BC$_2$N containing 16 valence electrons filling the VB. In the two binary compounds VB width amounts to 20 eV at Γ point center of the BZ, and ~22 eV for BC$_2$N due to a larger expansion of N s-like states. These characteristics are also similar to those of diamond.

In view of the iono-covalent character shown from the charge density projections, especially for *rh*-BC$_2$N, further illustration was sought from the site projected density of states (DOS) and the bonding as based the overlap integral S$_{ij}$ (i and j being two chemical species) within the COOP (crystal orbital overlap population) [33] implemented in the DFT-based augmented spherical wave ASW method [34]. Fig. 4a shows the DOS and Fig. 4b the integrated COOP translating the intensity/strength of the bonding. The zero energy along the *x*-axis is with respect to E$_V$, i.e. the top of the valence band. The energy gap is close to 5 eV as observed in Fig. 3. The VB shows a continuous shape albeit with two main blocks: (i) from -22 eV to -15 eV, the N(2s) states are dominant showing similarity with C2 and B s states; (ii) then p states show similar skyline for all constituents from -15 eV up to the top of VB with increasing intensity. The empty CB also shows mix of states arising from all constituents. The integrated COOP, *i*COOP panel shows significant results regarding the respective bonding strengths. At the bottom of the VB the non-directional s states show zero intensities; only significant *i*COOP are shown within the p block from -15 eV up to the top of the VB. Clearly, the inter-carbon C1-C2 and B-N interactions have the highest intensities and they ensure for the major bonding and large stability within the BC$_2$N structure. Less intensity is observed for C1-B and then C2-N with much less intensity. Both kinds of bonds are weakened by the large contribution of C1, B, C2 and N in the competitive C2-C1 and B-N interactions, respectively.

5- Mechanical properties.

*Bulk modules from the energy-volume equation of state*

From the ground state structures, a comparison of the respective energy-volume curves allows establishing the equations of state EOS. For the purpose, calculations were carried out at different volumes around minima found from the geometry optimization (cf. Table 2). The subsequent fit was done using the third-order Birch equation of state [22]:

$$E(V) = E_0(V_0) + \frac{9}{8}V_0 B_0[(V_0/V)^{2/3} - 1]^2 + \frac{9}{16}B_0(B'-4)V_0[(V_0/V)^{2/3} - 1]^3$$

where $E_0$, $V_0$, $B_0$, and $B'$ are the equilibrium energy, the volume, the bulk modulus, and its first pressure derivative, respectively.

Fig. 5a shows the $E(V)$ curves considering 2 atoms in 2D phases *rh*-BN *3R* proposed herein and *h*-BN *2H*, with the resulting fit values from above EOS. The two curves are overlapping with very close magnitudes of fit values. The $B_0$ magnitudes are characteristic of 2D graphite-like structures.

The same features are shown in Fig. 5b with the 3D *rh*-$B_2N_2$ compared to *c*-BN. The curves overlap, and the $B_0$ magnitude is much higher than in 2D phases. A slightly smaller magnitude for newly proposed *rh*-$B_2N_2$ is observed but the energy is almost the same for both and systematically smaller than for 2D phases.

Lastly, regarding 3D $BC_2N$, it was already shown in Table 1 that *rh*-$BC_2N$ has a larger stability with respect to orthorhombic (*o*-$BC_2N$) and cubic (*c*-$BC_2N$) structures. For the sake of comparison we elected to establish the EOS of *rh*-$BC_2N$ versus *o*-$BC_2N$. The corresponding $E(V)$ curves in Fig. 5c show a clear stabilization of *rh*-$BC_2N$ with 1 eV difference, albeit with similar volumes. The fit of the two curves with the EOS above provide the fit values in the inserts. The zero pressure bulk modulus, $B_0$ = 385 GPa is larger than for the two BN binaries in Fig. 5b. An assessment can be inferred from Fig. 4b which shows that both B-N and C-C bonding are prevailing in $BC_2N$, thus differentiating the ternary from the binary.

*Full elastic tensors*

We also obtained the elastic properties determined by performing finite distortions of the lattice and deriving the elastic constants from the strain-stress relationship. In hexagonal symmetry there are six independent elastic stiffness constants $C_{11}$, $C_{33}$, $C_{44}$, $C_{55}$, $C_{12}$, and $C_{13}$. Most encountered compounds are polycrystalline where monocrystalline grains are randomly oriented so that on a large scale, such materials can be considered as statistically isotropic. They are then completely described by the bulk modulus $B$ and the shear modulus $G$, which may be obtained by averaging the single-crystal elastic constants. The most widely used averaging method of the elastic stiffness constants is Voigt's one [28] based on a uniform strain. The calculated set of elastic constants in $B_2N_2$ and $BC_2N$ in units of GPa are:

$B_2N_2$: $C_{11}$ =908; $C_{12}$ = 140; $C_{13}$ = 50; $C_{33}$ = 952; $C_{44}$ = 385, and $C_{55}$ = 335.

$BC_2N$: $C_{11}$ = 1052; $C_{12}$ = 125; $C_{13}$ = 61; $C_{33}$ = 1058; $C_{44}$ = 467, and $C_{55}$ = 389.

All $C_{ij}$ values are positive and their combinations: $C_{11} > C_{12}$, $C_{11}C_{33} > C_{13}^2$ and $(C_{11}+C_{12})C_{33} > 2C_{13}^2$ obey the rules pertaining to the mechanical stability of the compound. The bulk ($B_V$) and shear ($G_V$) modules following Voigt are formulated as:

$$B_V = 1/9 \{2(C_{11} + C_{12}) + 4C_{13} + C_{33}\},$$

and

$$G_V = 1/30 \{C_{11} + C_{12} + 2 C_{33} - 4 C_{13} + 12 C_{44} + 6( C_{11} - C_{12})\}$$

The numerical values are then: $B_V(B_2N_2)$= 356 GPa; $G_V(B_2N_2)$ = 400 GPa, and $B_V(BC_2N)$= 412 GPa; $G_V(BC_2N)$ = 476 GPa. Although with different values, the values of $B_V$ follow the trends of magnitudes with the values obtained from the EOS fits (Fig. 5, inserts) thus validating the two different approaches. The Pugh's $G/B$ ratio [29] is an indicator of brittleness or ductility for $G/B > 0.5$ and $G/B < 0.5$, respectively. For $B_2N_2$ $G/B$ = 1.12, and $G/B$ = 1.15 for $BC_2N$, indicating high brittleness for both. This behavior could arise from directional bonding characteristics (*vide infra*).

Young's modulus ($E$) and Poisson's ratio (ν) were calculated from bulk and shear moduli using isotropic approximation (Table 3).

*Hardness*

Vickers hardness ($H_V$) of rhombohedral 3D $B_2N_2$ and $BC_2N$ was predicted using three contemporary theoretical models of hardness: (i) Mazhnik-Oganov model [23], (ii) Chen-Niu model [25], and (iii) Lyakhov-Oganov model [24]. The first two models use only the elastic properties, while Lyakhov-Oganov approach takes into account the strength of covalent bonding, degree of ionicity and directionality, and topology of the crystal structure. The results are presented in Table 3. For *rh*-$B_2N_2$ all three $H_V$-values are in surprising agreement, and are noticeably higher than Vickers hardness of single-crystal *c*-BN (62 GPa) but less than that of nano-crystalline *c*-BN (85 GPa) [30]. The average (from two models) Vickers hardness of rhombohedral BC$_2$N amounts to 87 GPa, which is higher than that of *c*-BC$_2$N (76 GPa [8]). One possible explanation can be found in the extended lattices in the predicted phases versus cubic ones.

Mazhnik-Oganov model was also used for estimation of fracture toughness ($K_{Ic}$) of (Table 3). For both of them $K_{Ic}$ is almost twice higher than the 2.8 MPa·m½ value for single-crystal *c*-BN [8] and close to fracture toughness values of *c*-BC$_2$N [8] and diamond [32].

6- Conclusions

Crystal chemistry rationale of structurally transforming 2D into 3D followed by geometry-optimization calculations within quantum density functional theory of the generated model structures, were applied to an important class of compounds of light elements B,C,N known for their generating ultra hard phases due to their small radii and their ability to embrace sp$^2$ (2D) and sp$^3$ (3D). New rhombohedral binary and ternary phases were predicted with exceptional mechanical properties based on energies and energy-derived qualities (charge density, equation of states, elastic constants,) backed with illustrations of the charge densities, the electronic band structures and the properties of chemical bonding. From the fact that the new 3D rhombohedral compounds are expressed in hexagonal settings, they can be considered as extended ultra-hard super-BN and super-BC$_2$N, i.e. 3D *h*-B$_6$N$_6$ and *h*-B$_3$C$_6$N$_3$.

**References**

[1] Lipson, H.; Stokes, A.R. A new structure of carbon. *Nature* 149 (1942) 328-956.

[2] Kurdyumov, A.V.; Solozhenko, V.L.; Zelyavski, W.B. Lattice parameters of boron nitride polymorphous modifications as a function of their crystal-structure perfection. *J. Appl. Cryst.* 28 (1995) 540-545.

[3] Wentorf, R.H. jr. Cubic form of boron nitride. *J.Chem. Phys.* 26 (1957) 956.

[4] Solozhenko, V.L. Häusermann, D.; Mezouar, M.; Kunz, M. Equation of state of wurtzitic boron nitride to 66 GPa. *Appl. Phys. Lett.* 72 (1998) 1691-1693.

[5] Solozhenko, V.L.; Chernyshev, V.V.; Fetisov, G.V.; Rybakov, V.B.; Petrusha, I.A. Structure analysis of the cubic boron nitride crystals. *J. Phys. Chem. Solids*, 51 (1990) 1011-1012.

[6] Solozhenko, V.L.; Andrault, D.; Fiquet, G.; Mezouar, M.; Rubie, D.C. Synthesis of superhard cubic $BC_2N$. *Appl. Phys. Lett.* 78 (2001) 1385-1387.

[7] Mattesini, M.; Matar, S.F. Search for ultra-hard materials: theoretical characterization of novel orthorhombic $BC_2N$ crystals. *Int. J. Inorg. Mater.* 3 (2001) 943–9572.

[8] Solozhenko, V.L.; Dub, S.N.; Novikov, N.N. Mechanical properties of cubic $BC_2N$, a new superhard phase. *Diam. Relat. Mater.* 10 (2001) 2228-2231.

[9] Hohenberg, P.; Kohn, W. Inhomogeneous electron gas. *Phys. Rev. B* 136 (1964) 864−871.

[10] Kohn, W.; Sham, L.J. Self-consistent equations including exchange and correlation effects. *Phys. Rev. A* 140 (1965) 1133−1138.

[11] Snis, A.; Matar, S.F. Electronic density of states, 1s core-level shifts, and core ionization energies of graphite, diamond, $C_3N_4$ phases, and graphitic $C_{11}N_4$. *Phys. Rev. B* 60 (1999) 10855-10863.

[12] Solozhenko, V.L.; Le Godec, Y. A hunt for ultrahard materials. *J. Appl. Phys.* 126 (2019) 230401.

[13] Kresse, G.; Furthmüller, J. Efficient iterative schemes for ab initio total-energy calculations using a plane-wave basis set. *Phys. Rev. B* 54 (1996) 11169-11186.

[14] Kresse, G.; Joubert, J. From ultrasoft pseudopotentials to the projector augmented wave. *Phys. Rev. B* 59 (1999) 1758−1775.

[15] Blöchl, P.E. Projector augmented wave method. *Phys. Rev. B* 50 (1994) 17953−17979.

[16] Perdew, J.; Burke, K.; Ernzerhof, M. The generalized gradient approximation made simple. *Phys. Rev. Lett*. 77 (1996) 3865−3868.

[17] Ceperley, D.M.; Alder, B.J. Ground state of the electron gas by a stochastic method. *Phys. Rev. Lett.* 45 (1980) 566-569.

[18] Press, W.H.; Flannery, B.P.; Teukolsky, S.A.; Vetterling, W.T. Numerical Recipes, 2nd ed. Cambridge University Press: New York, NY, USA, 1986.

[19] Blöchl, P.E.; Jepsen, O.; Anderson, O.K. Improved tetrahedron method for Brillouin-zone integrations. *Phys. Rev. B* 49 (1994) 16223-6.

[20] Methfessel, M.; Paxton, A.T. High-precision sampling for Brillouin-zone integration in metals. *Phys. Rev. B* 40 (1989) 3616-3621.

[21] Monkhorst, H.J.; Pack, J.D. Special k-points for Brillouin Zone integration. *Phys. Rev. B* 13 (1976) 5188-5192.

[22] Birch F. Finite strain isotherm and velocities for single-crystal and polycrystalline NaCl high pressures and 300 K. *J. Geophys. Res*. 83 (1978) 1257-1259.

[23] Mazhnik, E.; Oganov, A.R. A model of hardness and fracture toughness of solids. *J. Appl. Phys.* 126 (2019) 125109.

[24] Oganov, A.R.; Lyakov, A.O. Evolutionary search for superhard materials: Methodology and applications to forms of carbon and $TiO_2$. *Phys. Rev. B* 84 (2011) 092103.

[25] Chen, X-Q; Niu, H.; Li D. Modeling hardness of polycrystalline materials and bulk metallic glasses. *Intermetallics* 19 (2011) 1275-1281.

[26] Matar, S.F; Pöttgen, R. Coloring in the ZrBeSi type structure. *Z. Naturforsch. B* 74 (2019) 307-318.

[27] Momma, K.; Izumi, F. VESTA 3 for three-dimensional visualization of crystal, volumetric and morphology data. *J. Appl. Cryst*. 44 (2011) 1272–1276.

[28] Voigt, W. Über die Beziehung zwischen den beiden Elasticitätsconstanten isotroper Körper. *Annalen Phys.* 274 (1889) 573–587.

[29] S. F. Pugh Relations between the elastic moduli and the plastic properties of polycrystalline pure metals. *Phil. Mag*. 45 (1954) 823-843.

[30] Solozhenko, V.L.; Bushlya, V.; Zhou, J. Mechanical properties of ultra-hard nanocrystalline cubic boron nitride. *J. Appl. Phys.* 126 (2019) 075107.

[31] Sun, H.; Jhi, S.-H.; Roundy, D.; Cohen, M.L.; Louie, S.G. Structural forms of cubic $BC_2N$. *Phys. Rev. B* 64 (2001) 094108-094108.6

[32] Novikov, N.V.;, Dub, S.N. Fracture toughness of diamond single crystals. *J. Hard Mater.* 2 (1991) 3-11.

[33] Hoffmann, R. How chemistry and physics meet in the solid state. *Angew. Chem. Int. Ed.* 26 (1987) 846–878.

[34] Eyert, V. Basic notions, and applications of the augmented spherical wave ASW method. *Int. J. Quant. Chem.* 77 (2000) 1007–1031.

TABLES

Table 1   Total and atom averaged energies of the different phases considered in this work (*pw*: present work).

| Phase | Space Group | $E_{Tot.}$ (eV) | E/at. (eV) |
|---|---|---|---|
| 2D binary B-N | | | |
| *h*-BN (*2H*) (*pw*) | P6$_3$/mmc N°194 | -17.54 | -8.77 |
| *rh*-B$_2$N$_2$ (*3R*) (*pw*) | R3m N°160 | -17.54 | -8.77 |
| | | | |
| 3D binary B-N | | | |
| *c*-BN [2] | F$\bar{4}$3m N°216 | -17.42 | -8.71 |
| *rh*-B$_2$N$_2$ (*pw*) | R3m N°160 | -34.84 | -8.71 |
| | | | |
| 3D ternary B-C-N | | | |
| *rh*-BC$_2$N (*pw*) | R3m N°160 | -34.87 | -8.72 |
| *c*-BC$_2$N [31] | P1 N°1 | -67.81 | -8.47 |
| *o*-BC$_2$N [7] | Pmm2 N°25 | -33.87 | -8.46 |

Table 2   Calculated structure parameters. In the rhombohedral cell, all atoms are at 1*a* Wyckoff position (*x, x, x*) or in 3*a* position (0, 0, *z*) in hexagonal settings, *x* and *z* are the same in both settings.

a) Rhombohedral 3*R* BN, space group *R*3*m* (N°160). $a_{rh}$= 3.51 Å; α = 41.8°.
Parameters in hexagonal setting. $a_{hex}$ = 2.502 Å; $c_{hex}$ = 9.595 Å. i.e. $h$-$B_2N_2$

| Atom | Wyckoff | x | y | z |
|---|---|---|---|---|
| B | 3*a* | 0.0 | 0.0 | 0.166 |
| N | 3*a* | 0.0 | 0.0 | 0.834 |

d(B-N) =1.45 Å.

b) Rhombohedral 3D $B_2N_2$, space group *R*3*m* (N°160). $a_{rh}$= 4.44 Å; α = 33.56°.
Crystal parameters in hexagonal setting, i.e. $h$-$B_3C_6N_3$ with $a_{hex}$= 2.56 Å; $c_{hex}$ 12.56 Å.

| Atom | x | y | z |
|---|---|---|---|
| B1 | 0.0 | 0.0 | 0.188 |
| B2 | 0.0 | 0.0 | 0.688 |
| N1 | 0.0 | 0.0 | 0.312 |
| N2 | 0.0 | 0.0 | 0.812 |

d(B1-N2)= 1.56 Å, d(B2-N1)= 1.57 Å

ϕ(B-N1-B1) = ϕ(N-B1-N1)=109.00°

ϕ(N1-B1-N1) = ϕ(B1-N1-B1)=109.94°

c) Rhombohedral 3D $BC_2N$, space group *R*3*m*, N°160. $a_{rh}$= 4.45 Å; α = 33.14°.
Crystal parameters in hexagonal setting, with $a_{hex}$= 2.54 Å; $c_{hex}$ 12.56 Å. i.e. for $B_3C_6N_3$ stoichiometry
In rhombohedral cell, all atoms are at 1*a* Wyckoff position *x, x, x* or in 3*a* position 0, 0, *z* in hexagonal setting.

| Atom | x | y | z |
|---|---|---|---|
| C1 | 0 | 0 | 0.184 |
| C2 | 0 | 0 | 0.811 |
| B  | 0 | 0 | 0.314 |
| N  | 0 | 0 | 0.690 |

d(C2-N)= 1.52 Å, d(C1-C2)= 1.55 Å, ,d(B-N)= 1.57 Å, d(C1-B)= 1.63 Å

ϕ(C2-C1-C2) = 110.09°, ϕ(N-B-N) = ϕ(B-N-B) 108.78°,

ϕ(C2-C1-B) = 108.84° ϕ(B-N-B) = ϕ(N-B-N)=109.00°  ϕ (C1-C2-C1) = 110.09°

Table 3 Predicted mechanical properties of rhombohedral 3D $B_2N_2$ and 3D $BC_2N$: Vickers hardness ($H_V$), bulk modulus ($B_0$), shear modulus ($G$), Young's modulus ($E$), Poisson's ratio ($\nu$) and fracture toughness ($K_{Ic}$).

|  | $H_V$ | | | $B_0$ | $G$ | $E^§$ | $\nu^§$ | $K_{Ic}^*$ |
|---|---|---|---|---|---|---|---|---|
|  | MO[*] | CN[†] | LO[‡] |  |  |  |  |  |
|  | GPa | | | | | | | MPa·m$^{½}$ |
| $rh$-$B_2N_2$ | 75 | 73 | 74 | 356 | 400 | 873 | 0.091 | 4.4 |
| $rh$-$BC_2N$ | 90 | 84 | – | 412 | 476 | 1031 | 0.083 | 5.7 |

[*] Mazhnik-Oganov model [23]

[†] Chen-Niu model [25]

[‡] Lyakhov-Oganov model [24]

[§] $E$ and $\nu$ values calculated using isotropic approximation.

**F I G U R E S**

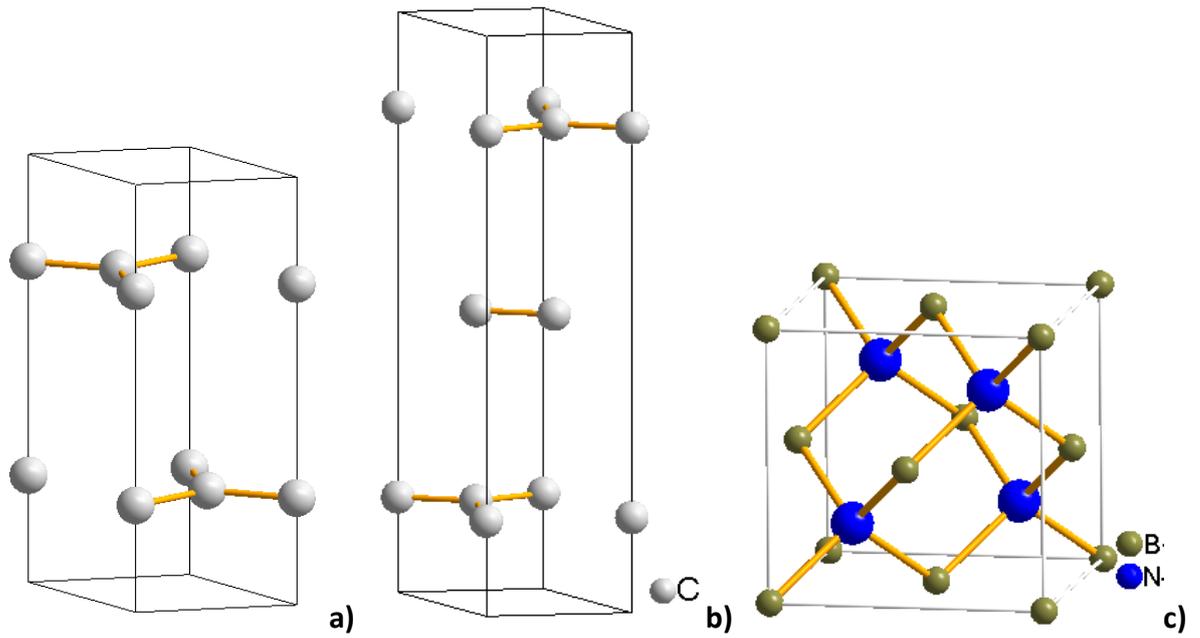

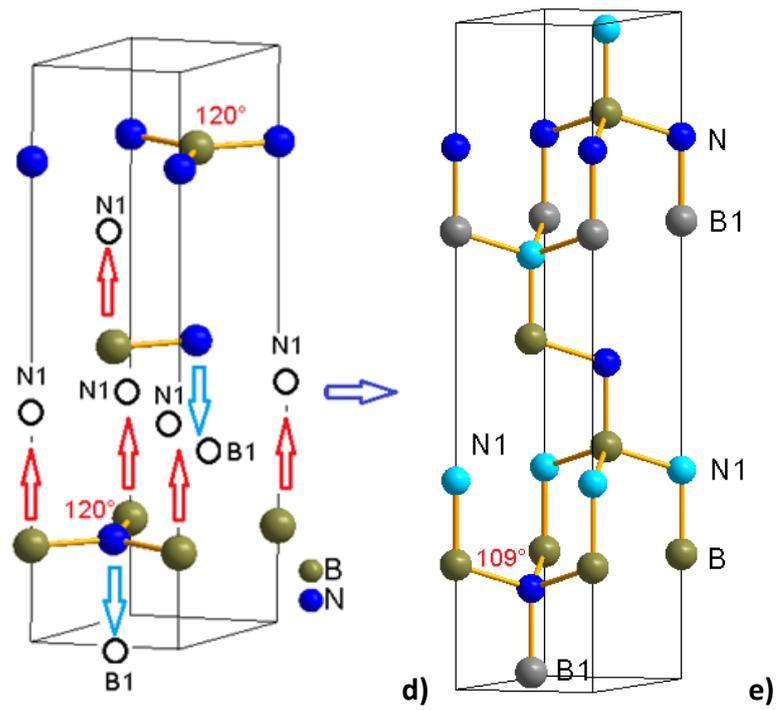

**(continued)**

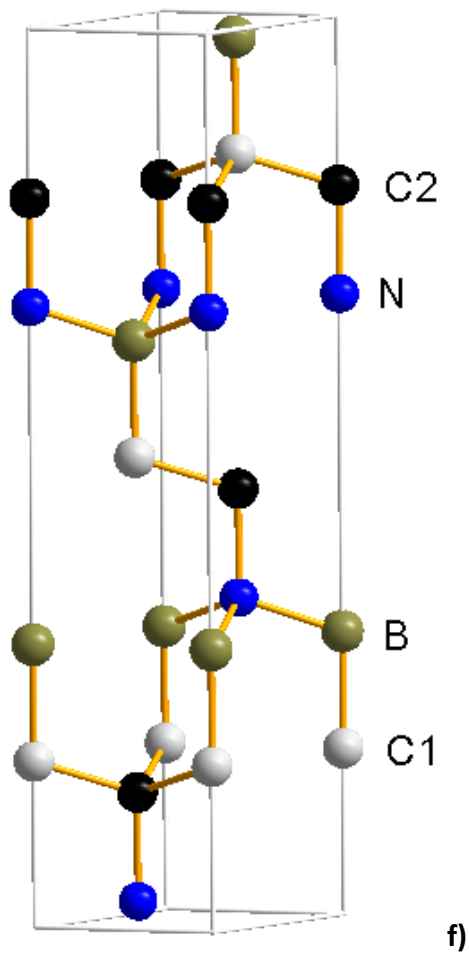

Fig. 1 Crystal structures: a) $2H$ $C_4$; b) $3R$ $C_2$ or ($h$-$C_6$); c) $c$-BN; d) 2D → 3D transformation mechanism; e) $rh$-$B_2N_2$; f) $rh$-$BC_2N$. All rhombohedral structures are expressed in hexagonal settings, i.e. with three times more atoms (2D $h$-$B_3N_3$, and 3D $h$-$B_6N_6$ and $h$-$B_3C_6N_3$).

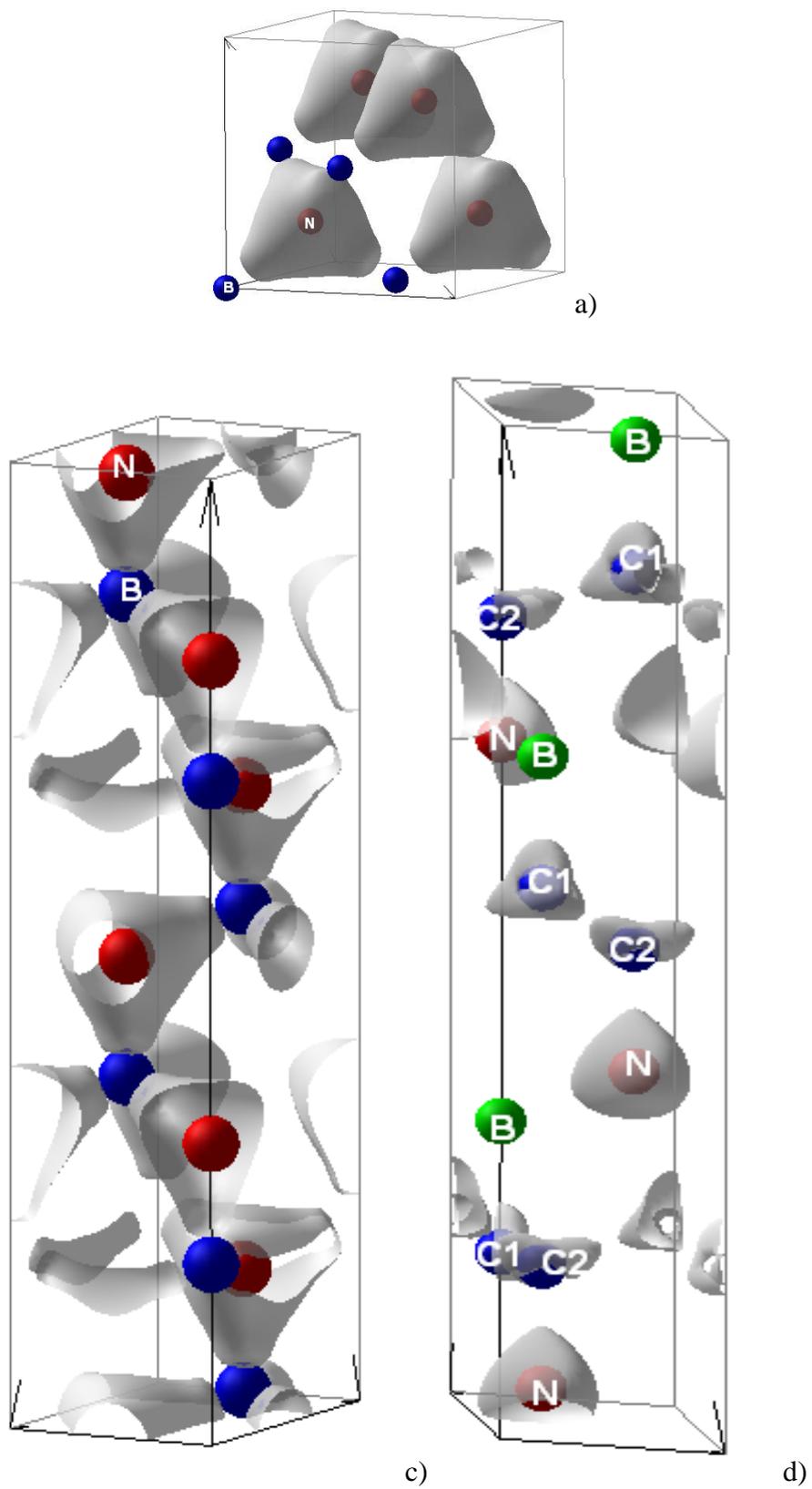

Fig. 2  Charge density isosurfaces of 3D phases: a) *c*-BN,  c) *h*-B$_6$N$_6$ and d) *h*-B$_3$C$_6$N$_3$.

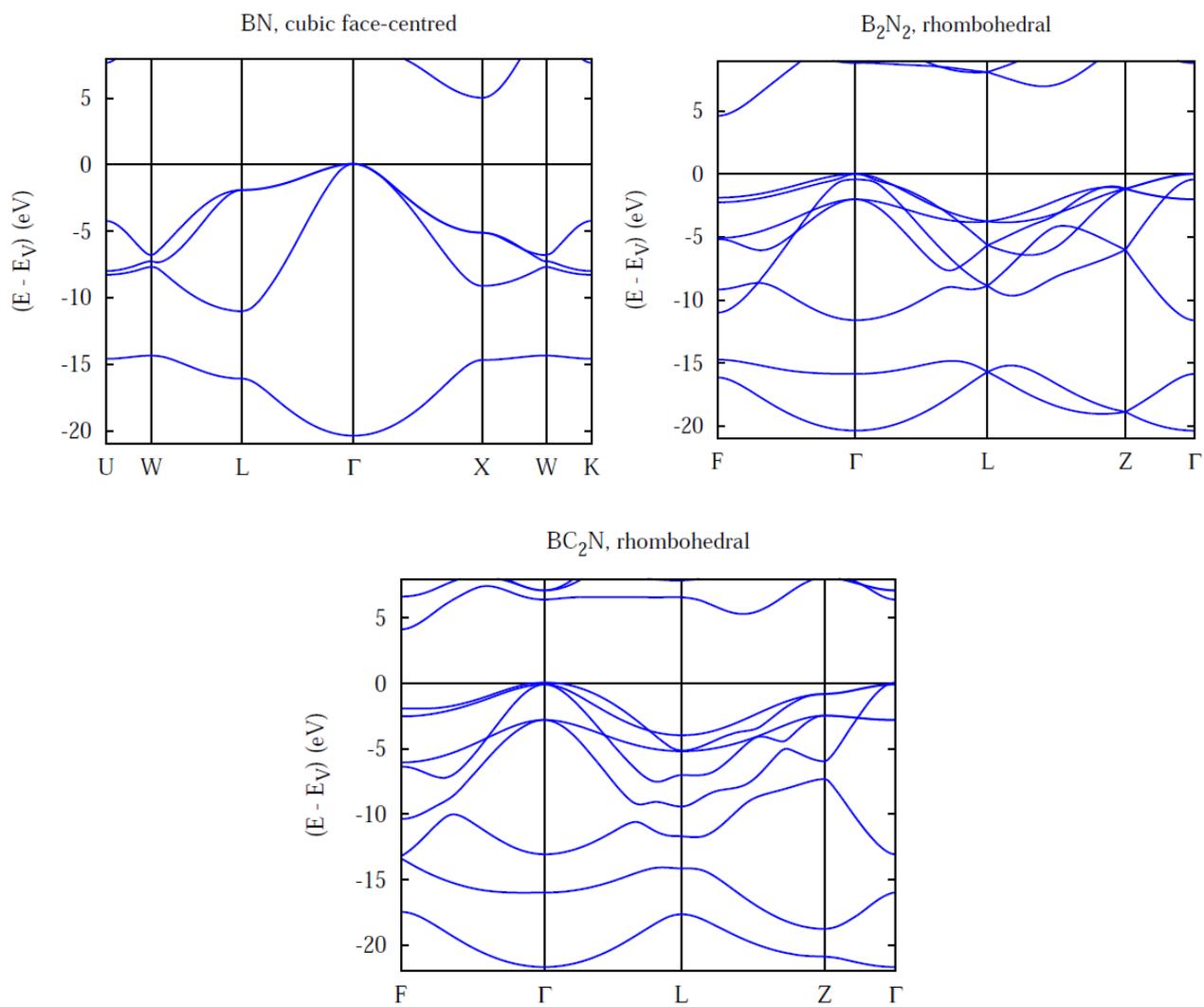

Fig. 3 Electronic band structures

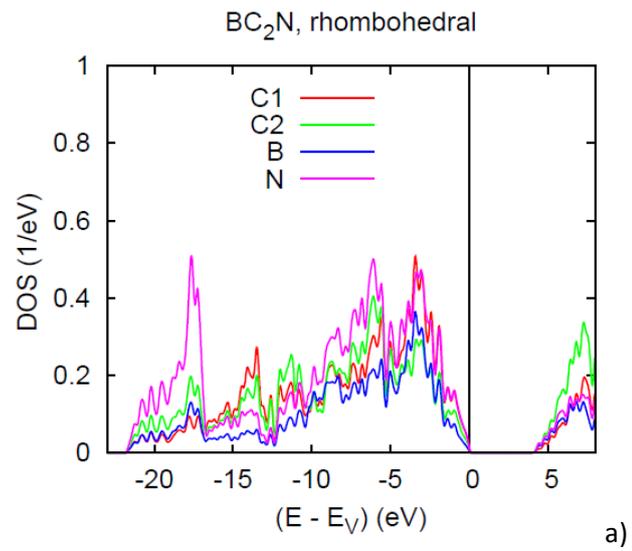

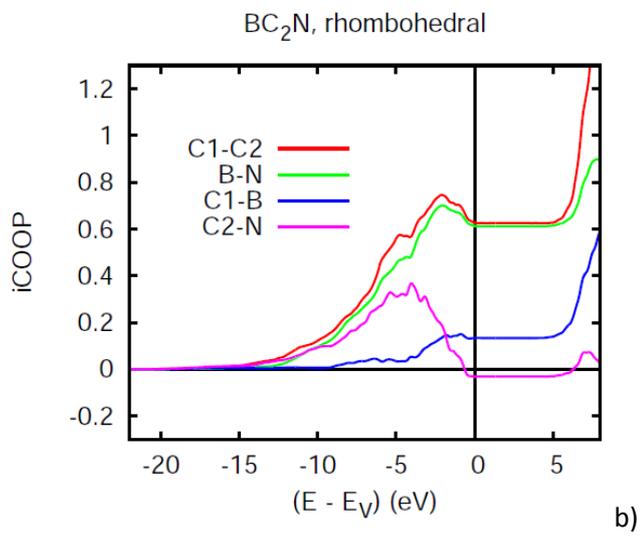

Fig. 4 BC$_2$N: a) Electronic density of states, b) Bonding from integrated COOP

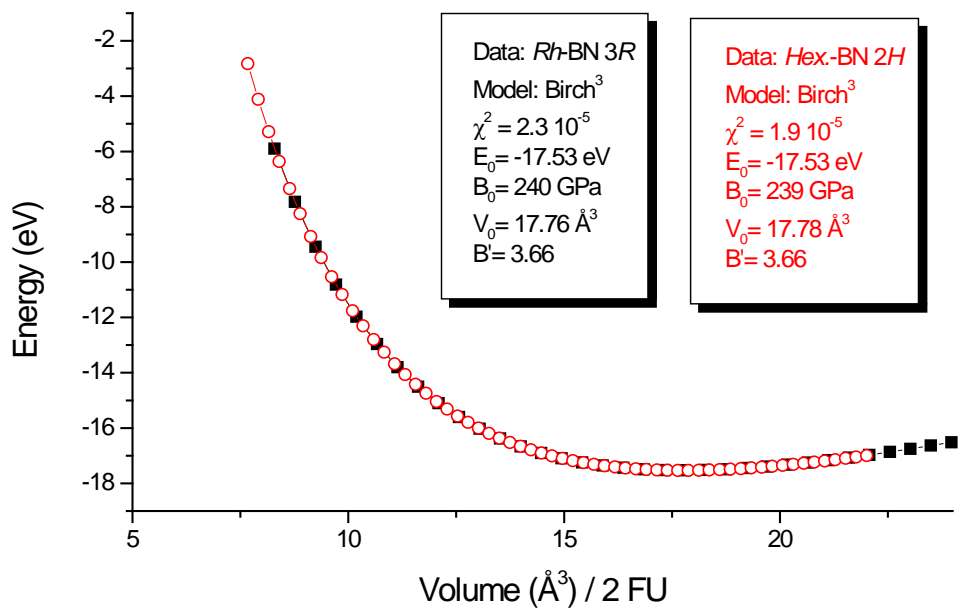

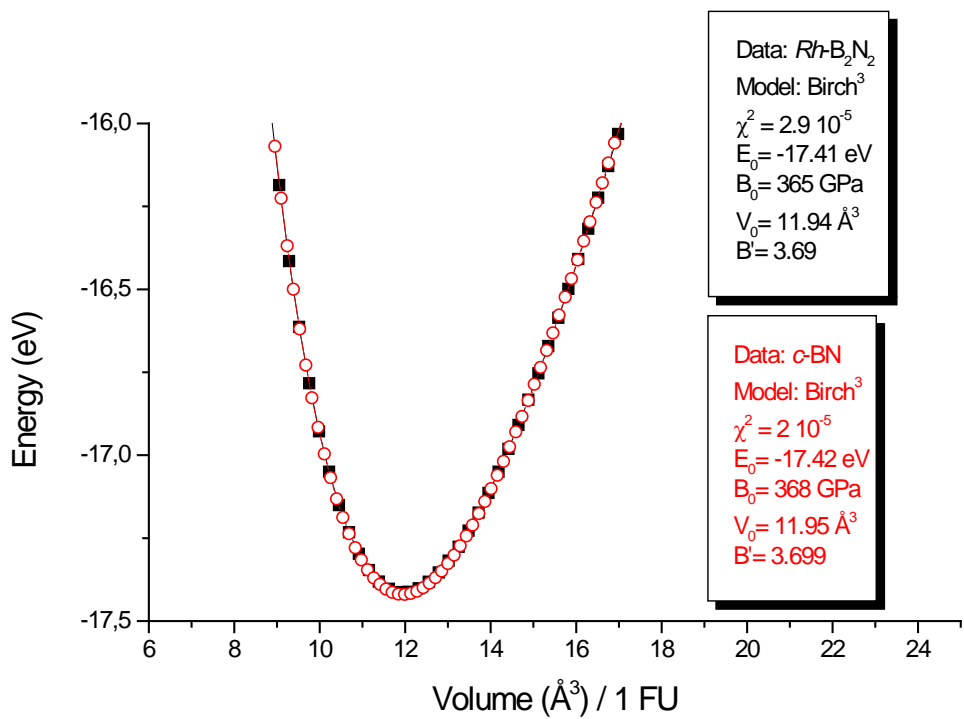

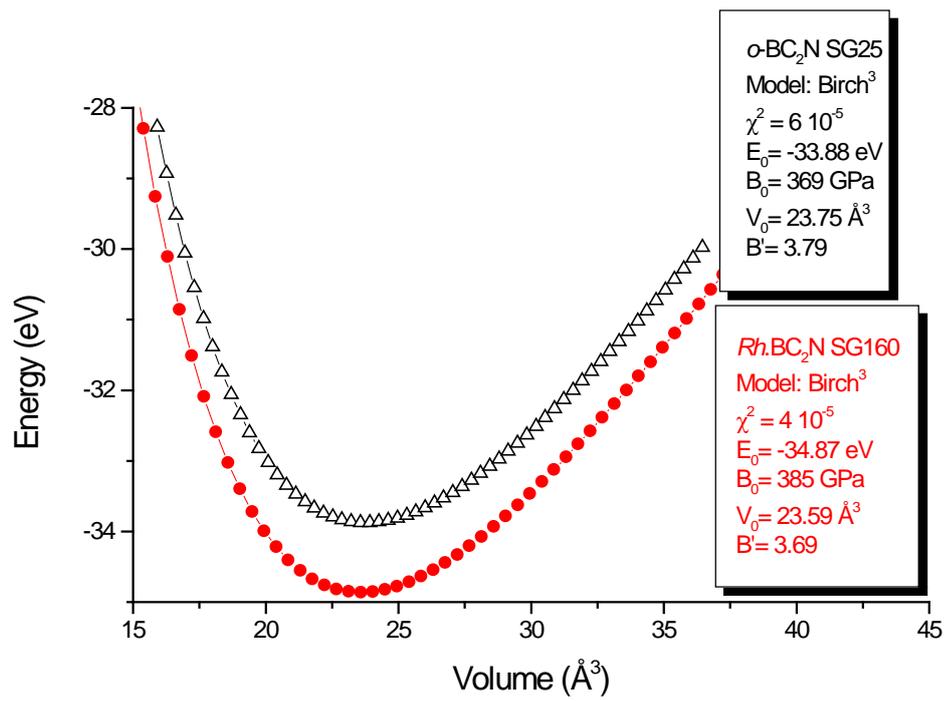

Fig. 5 Energy-volume curves and 3$^{rd}$-order Birch EOS fit values